\documentclass[prx,twocolumn,showpacs,floatfix,superscriptaddress]{revtex4-1}

\usepackage{amssymb,amsmath,amstext}                
\usepackage{graphicx}                                               
\usepackage{epstopdf}                                               
\usepackage{color}                                                     
\usepackage{bm}                                                        
\usepackage{appendix}                                              
\usepackage[utf8]{inputenc}
\usepackage{bbold}
\usepackage{bbm}
\usepackage{latexsym}
\usepackage[colorlinks=true,citecolor=blue,linkcolor=magenta]{hyperref}

\begin{document}

\title{At the limits of criticality-based quantum metrology:\\
apparent super-Heisenberg scaling revisited}

\author{Marek M. Rams}
\affiliation{Instytut Fizyki im. Mariana Smoluchowskiego, Uniwersytet Jagiello\'nski,  \L{}ojasiewicza 11, 30-348 Krak\'ow, Poland }
\email{marek.rams@uj.edu.pl}

\author{Piotr Sierant}
\affiliation{Instytut Fizyki im. Mariana Smoluchowskiego, Uniwersytet Jagiello\'nski,  \L{}ojasiewicza 11, 30-348 Krak\'ow, Poland }
\email{piotr.sierant@uj.edu.pl}

\author{Omyoti Dutta}
\affiliation{Instytut Fizyki im. Mariana Smoluchowskiego, Uniwersytet Jagiello\'nski,  \L{}ojasiewicza 11, 30-348 Krak\'ow, Poland }
\affiliation{Donostia International Physics Center DIPC, Paseo Manuel de Lardizabal 4, 20018 Donostia-San Sebasti\'{a}n, Spain.}

\author{Pawe\l{} Horodecki}
\affiliation{Faculty of Applied Physics and Mathematics, Gda\'nsk University of Technology,
ulica Gabriela Narutowicza 11/12, 80-233 Gda\'nsk, Poland}
\affiliation{National Quantum Information Center of Gda\'nsk, ulica W\l{}adys\l{}awa Andersa 27, 81-824 Sopot, Poland}
\email{pawel@mif.pg.gda.pl}

\author{Jakub Zakrzewski}
\affiliation{Instytut Fizyki im. Mariana Smoluchowskiego, Uniwersytet Jagiello\'nski,  \L{}ojasiewicza 11, 30-348 Krak\'ow, Poland }
\affiliation{Mark Kac Complex
Systems Research Center, Uniwersytet Jagiello\'nski, Krak\'ow,
Poland. }
\email{jakub.zakrzewski@uj.edu.pl}

\date{\today}

\begin{abstract}
We address the question whether the super-Heisenberg scaling for quantum estimation is indeed realizable. 
We unify the  results of two approaches.
In the first one,  the original system is compared with its copy rotated by the parameter dependent 
dynamics. If the parameter is coupled to the 
one-body part of the Hamiltonian  the precision of its estimation is known to scale at most as $N^{-1}$ (Heisenberg scaling)
in terms of the number of elementary subsystems used, $N$.  The second approach compares the overlap 
between the ground states of the parameter dependent Hamiltonian in critical systems, often leading to an apparent super-Heisenberg scaling.
However, we point out that if one takes into account the
scaling of time needed to perform the necessary operations, i.e. ensuring adiabaticity of the evolution, the Heisenberg limit given by the rotation scenario is recovered.
We illustrate the general theory  on a ferromagnetic Heisenberg spin chain  example
and show that it exhibits such super-Heisenberg scaling of ground state fidelity around the critical value of the parameter (magnetic field) governing the one-body part of the Hamiltonian.
Even an elementary estimator represented by a single-site magnetization  already outperforms the Heisenberg behavior  providing the $N^{-1.5}$ scaling. In this case Fisher information sets the ultimate scaling as $N^{-1.75}$ which can be saturated by measuring magnetization on all sites simultaneously. We discuss universal scaling predictions of the estimation precision offered by such observables, both at zero and finite temperatures, and support them with numerical simulations in the model. We provide an experimental proposal of realization of the considered model  via mapping the system to  ultra-cold bosons in periodically shaken optical lattice.  We explicitly derive that the Heisenberg limit
is recovered when time needed for preparation of quantum states involved is taken into acocunt.
\end{abstract}

\maketitle

\section{Introduction}
At the center of quantum metrology  \cite{Caves81,Wineland92,BraunsteinCaves94,Giovannetti04} lays the concept of estimation of a small external parameter with 
the help of a quantum procedure. The main idea is to engineer a family of quantum states depending strongly 
on that parameter in the sense that a small difference in the parameter value makes the states
 significantly different from each other. 
The relevant quantifier of a distance between quantum states is the quantum fidelity \cite{Uhlmann76}
\begin{equation} 
{\cal F}(\hat  \rho, \hat \sigma)=\mathrm{Tr}\left(\sqrt{\sqrt{\hat \rho} \hat \sigma \sqrt{\hat \rho}}\right),
\label{eq:fidelity}
\end{equation}
where density operators $\hat \rho$ and $\hat \sigma$ describe the states being compared. 

Now consider a family of quantum states $\hat \rho(\lambda)$  controlled by a parameter $\lambda$ and let $\delta_\lambda$ 
be a small shift of the parameter that we want to estimate. An ultimate bound on the accuracy of any estimate one may make on 
the unknown $\delta_\lambda$ is set be the quantum Fisher information (QFI) \cite{BraunsteinCaves94,Taddei_Quantum_speed_limit_2013}
\begin{equation}
G(\lambda) = - 4 \left. \partial^2 {\cal F}(\hat \rho (\lambda) , \hat \rho(\lambda + \delta_\lambda))/ \partial{\delta_\lambda}^2 \right \vert_{\delta_{\lambda}=0}.
\label{eq:QFI_definition}
\end{equation}
Indeed, in order to identify $\delta_\lambda$ one has to measure some observable $\hat A$, called an estimator. The precision it offers  
is quantified by the error propagation formula given by the inverse of signal-to-noise ratio
\begin{equation}
\Delta_{\delta_\lambda} (\hat A,\lambda) = \frac{\sqrt{\langle \hat A^{2} \rangle_{\rho(\lambda)} - \langle \hat A \rangle_{\rho(\lambda)}^{2}}}{ \left| \left. \frac{\partial \langle \hat A \rangle_{\rho(\lambda+ \delta_\lambda)}}{\partial \delta_\lambda} \right|_{\delta_\lambda =0} \right|}.
\label{eq:epf}
\end{equation}
The ultimate lower bound for the uncertainty 
of estimation of  {the}  small deviation $\delta_\lambda$ is set by the quantum 
Cramer-Rao bound \cite{BraunsteinCaves94},
\begin{equation}
\Delta_{\delta_\lambda} (\hat A, \lambda) \geq {G(\lambda)}^{-1/2},
\label{eq:C-Rao}
\end{equation}
that is independent of the observable $\hat A$ and determined by the QFI.
In principle, the above bound would be saturated by some special observable $\hat {\tilde {A}}$,  called symmetric logarithmic derivative operator, satisfying $2 \frac{\partial \hat \rho(\lambda) } {\partial \lambda} = \hat{\tilde{A}}  \hat  \rho(\lambda) +  \hat \rho(\lambda) \hat{\tilde{A}}$.
In practice, however,  an identification of an appropriate symmetric logarithmic derivative operator is a formidable task in itself. We shall show that for a quite general class of systems a possible alternative yielding in some limits at least a correct scaling with the system size exists.

There are  basically two different scenarios discussed in the literature on how to introduce 
the dependence of the state on the parameter $\delta_\lambda$.  In the first approach 
\cite{Giovannetti2011,Toth14,Demkowicz-review}
the  state is rotated by some Hamiltonian and then the estimator observable $\hat A$ is 
measured -- providing the accuracy that is determined by the error propagation formula. Let us call it a rotation scenario.

In principle, with many-body interacting Hamiltonian the 
corresponding Fisher information could have implied the error vanishing exponentially  with $N$ \cite{RoyBraunstein08}. 
It has been proven, however, that when the Hamiltonian is composed only of local on-site (or one-body - see e.g. \cite{boixo07}) terms $\hat h_{n}$, i.e.
\begin{equation}
\hat  H=\lambda \hat H_1 = \lambda \sum_{n=1}^{N} \hat h_{n},
\label{aconiechmaswoje}
\end{equation}
 then at most $G^{-1/2} \sim N^{-1}$. Such a scaling is referred to as the Heisenberg limit and should be contrasted with the classical type of behavior where $G^{-1/2}\sim N^{-1/2}$, i.e. the shot noise limit. It has been argued that adding to the above Hamiltonian other interactions -- not coupled to $\lambda$ --  cannot improve the scaling beyond the Heisenberg limit \cite{Giovanetti06,boixo07,Pasquale2013,Dur, Pang_2014,Fraisse2016}.
Furthermore, let us mention that the final formula is quite sensitive to a local noise and because of that one basically always ends up with the classical scaling  for  large enough $N$ \cite{Demkowicz-problem}. 

More precisely, bringing the time of the evolution explicitly into the picture,  for the Hamiltonian of the form
 \begin{equation}\hat  H=  \hat H_0 +\lambda  \hat H_1,
 \label{hamsite2}
 \end{equation}
it has been proven that  \cite{boixo07},
\begin{equation}
G^{-1/2} \ge \frac{1}{t ||\hat H_1||}.
\label{eq:boixo}
\end{equation}
 Above, $t$ is the time of the evolution and $||\hat H_1||$ is the norm of the operator coupled to $\delta\lambda$. 
 Most importantly, as the time factor might be experimentally limited, the focus usually is on the scaling of the norm only.  
 {The above bound holds for any initial state. Saturating it, if at all possible in the general case (even only in the limit of short times), usually requires considering highly entangled GHZ-like probe states.}
 
When the Hamiltonian $\hat H_1$ above includes $k$-body terms, the possible scaling shifts to $G^{-1/2} \sim N^{-k}$ \cite{boixo07}, provided that all possible $k$-body subsets are present in $\hat H_1$ to contribute to the norm in Eq.~\eqref{eq:boixo}. In principle,  this might allow to go beyond the Heisenberg limit for $k\ge2$. Such democratic couplings are, however, difficult  to create in Nature.
Recently, those results were extended to describe both open and noisy systems \cite{Alipour2014,Alipour2015,adolfo1,adolfo2,Demkowicz17}.  

The rotation scenario serves also as a powerful entanglement detector \cite{Laskowski,Toth12}
which can detect a vanishing  fraction of entanglement \cite{Augusiak}
or even bound entanglement \cite{Czekaj}  with scaling close to the Heisenberg limit. Entanglement detection method has recently found a new application in the proposal 
to extract the Fisher information from a dynamical susceptibility of the thermal input state  \cite{Hauke16} (see also \cite{Apellaniz15}), which is measurable, for instance, by means of Bragg spectroscopy \cite{Ernst2010,Inguscio}. Recently, the pure state metrology in the spirit of the rotation scenario has been reformulated in terms of the Loschmidt echo \cite{Tsang13,Loschmidt}.   Let us also mention that one can consider Fisher information as a detector of 
non-equilibrium phase transitions \cite{ugo1,ugo2} or for multipartite entanglement questions \cite{Raja2017}. For a review of quantum enhanced measurements without entanglement see \cite{braun17}.

Having all the above in mind, one immediately recalls the second approach that connects estimation problem to the concept of 
criticality \cite{Zanardi2006,You-Li-Gu,Paris,Invernizzi08,Salvatori2014,Bina2016,Kraus16,sanpera_2016,Frerot17}.
In that approach one focuses on the situation where the dependence of the state on $\lambda$ has a
completely different origin. The state $|\Psi(\lambda)\rangle$  is the ground state of 
the Hamiltonian depending on the parameter $\lambda$, which exhibits  criticality at some critical  point  $\lambda_{c}$.  
The essence of this approach is an observation that in the vicinity of the critical point the ground states 
overlap becomes drastically sensitive to small change of  $\lambda$. Clearly, this sensitivity is again quantified by QFI.

In the context of ground state fidelity (or more generally for the thermal states) it is customary to introduce fidelity susceptibility $\chi_F(\lambda)$.  For sufficiently small $\delta_\lambda$, in a finite system, one has
\begin{equation}
{\cal F}(\hat \rho (\lambda) , \hat \rho(\lambda + \delta_\lambda))= 1 - \frac{1}{2} \chi_F (\lambda) \delta_\lambda^{2} + O( \delta_\lambda^{3}).
\label{eq:rozwiniecie}
\end{equation}
Fidelity susceptibility and QFI are directly proportional to the Bures distance between  density matrices at slightly differing values of $\lambda$ \cite{Hubner93, Invernizzi08} and $G(\lambda)=4  \chi_F (\lambda) $.

Interestingly, it has been observed in this case that  for Hamiltonian \eqref{hamsite2}, criticality can boost QFI to $G^{-1/2} \sim N^{-l}$ with $ 2 < l < 3$, see \cite{Gong-Tong08,Gu-Kwok-Ning08,Greschner-Kolezhuk-Vekua13}, leading to an apparent super-Heisenberg scaling. There seems to exist a clear contradiction with the rotation scenario. Can these two pictures be reconciled? This is the main aim of the present work --  we solve this super-Heisenberg puzzle. We show that the overlap measurement contains an additional ingredient, namely the time it takes to transform one ground state into another one at a slightly different parameter value.  This may be translated into the additional $N$-power scaling of time in the vicinity of the critical point if one assumes adiabatic dynamics, which is a necessity if we are to compare ground states.  This allows us to reconcile the two approaches to quantum metrology.

The rest of the article is organized as follows.
In Sec.~\ref{Sec_basic} we define and discuss basic properties of fidelity and fidelity susceptibility, the main tools of the analysis that follow. 
Sec.~\ref{main} contains the main results of our work. Using finite size scaling hypothesis based on the renormalisation group approach, a well established tool in the treatment of quantum criticality, we 
derive the scaling of precision offered by the most natural observables coupled to the perturbation. In the adiabatic limit they can saturate the ultimate scaling set by QFI.
Most importantly, we also bring the time directly into the picture and discuss the appropriate time scale necessary to recover the adiabatic dynamics at the critical point and reach the above-mentioned scalings.
By factoring out the evolution time we illustrate that the ground state approach naturally satisfies the Heisenberg limit as it is understood in the rotational scenario. The apparent super-Heisenberg scalings are recovered in the limit of sufficiently long evolution times and can be understood as the ultimate limit of precision which can be obtained within this approach (whatever the time is). Using a critical ground state as a probe state, nevertheless, might allow to break the shot-noise limit due to strong correlations/entanglement in such a state.

The general theory is illustrated on a particular example in Sec.~\ref{Sec_XYmodel}. We discuss the ferromagnetic Heisenberg spin chain where the parameter to be estimated is a small external magnetic field.  This model provides a minimal entanglement model in a sense that $H_1$ is separable while $H_0$ involves only 2-body (nearest neighbors) terms. Here we test the universal scaling of the error propagation formula for those natural observables against numerical data in the immediate vicinity of the critical point of the model. In particular,  we obtain in this model $G^{-1/2}(\lambda_c) \sim N^{-1.75}$ with $\lambda_c$ being the parameter value at the critical point. Moreover, unlike in the standard, rotation-based  metrology, the most natural, strictly local and parameter independent estimator, namely the single-site magnetization, is enough to go beyond the apparent Heisenberg limit by reaching $N^{-1.5}$.  We also show that the operator which measures the magnetization on all sites simultaneously 
scales in the same way as the optimal one. We should point out that finding an analytical form of the optimal operator in many-body system is typically a daunting challenge, see e.g. \cite{Paris,Invernizzi08}. With all these interesting properties, when 
a time factor is properly taken into account (i.e. the time needed to adiabatically transfer a ground state into another ground state at different value of the parameter) we recover the Heisenberg limit. 

The possible realization of this model  in cold atom optical lattice setting is  given in Sec.~\ref{sec:cold}. We found it appropriate to first extend the discussion from smooth adiabatic quench to the instantaneous one, i.e. the Loschmidt echo, arguing that similar universal behavior can be observed also in that case. This is discussed in Sec.~\ref{sec:le}. In Sec.~\ref{sec:detune} we consider the robustness of the observed features, i.e. we consider the  situation detuned from criticality as well as the impact of finite temperature. We conclude in Sec.~\ref{sec:discuss}.
Finally, in the Appendix~\ref{sec:swap} we discuss an universal estimator-type measurement in the paradigm where the original reference state and the specific quadratic interactions are accessible.

\section{Basics of fidelity susceptibility}
\label{Sec_basic}

Consider the quantum system   depending on a parameter $\lambda$, the value of which we shall
try to estimate. For the ground state $|\Psi(\lambda)\rangle$ of the Hamiltonian $\hat H(\lambda)$
the fidelity defined in Eq.~\eqref{eq:fidelity} simplifies as
\begin{equation}
{\cal F}= |\langle \Psi(\lambda)|\Psi(\lambda+\delta_\lambda)\rangle|.
\label{eq:def_F}
\end{equation}
It is intuitively clear that fidelity may be significantly below unity, or alternatively that $\chi_F(\lambda)$ (compare Eq.~\eqref{eq:rozwiniecie}) is large, when the properties of the 
system change significantly with $\lambda$. Then the measurement of some observable might lead to an accurate 
determination of $\lambda$. Clearly, when the system undergoes the quantum phase transition its properties change 
dramatically, that is why the maxima of $\chi_F(\lambda)$ (for a finite system) or its divergences (in the thermodynamic limit) signal the 
location of the quantum critical point \cite{Zanardi2006}. Obviously, for Eq.~\eqref{eq:rozwiniecie} to hold we have to consider a finite system and sufficiently small $\delta_\lambda$ -- otherwise  higher terms in that expansion are non-negligible and one should be considering $\log \mathcal{F}$ which becomes an extensive quantity in that limit \cite{Zhou2008,MMR2011a,*MMR2011b}.  

It has been shown that the universal information can be extracted from the behavior of fidelity susceptibility in the vicinity of the critical point  \cite{zanardi_geometric,Schwandt2009,ABQ2010,PolkovnikovArXiv2010}.
To that end, and in order to relate directly to the rotational scenario in Eqs.~(\ref{aconiechmaswoje}--\ref{eq:boixo}), we consider the Hamiltonian
 \begin{equation}
 \hat  H(\lambda)=  \hat H_0 + \lambda  \hat H_1 = \hat H_0 + \lambda \sum_n{ \hat h_n},
 \label{ciacho}
 \end{equation}
specifying it to be in a broad class of systems consisting of $N = L^d$ spins in $d$ spatial dimensions which has a continuous critical point at $\lambda_c$. The general concept and scaling analysis \cite{Sachdev,Shondireview,ContinentinoBook} naturally applies also to systems of fermions and bosons.
The perturbation coupled to $\lambda$ in Eq.~\eqref{ciacho} consists of local on-site terms, note however that the same would hold for $\hat h_n$ having support on a couple of neighboring sites.
$\hat H_1$ is a relevant renormalisation group perturbation which drives the transition and we assume that it has a well defined scaling dimension.
The divergence of the correlation length in the vicinity of the critical point, $\xi \sim |\lambda-\lambda_c|^{-\nu}$, specifies the critical exponent $\nu$.

The universal part of the fidelity susceptibility  at the critical point is expected to scale as  \cite{Schwandt2009,ABQ2010,PolkovnikovArXiv2010} 
\begin{equation}
G(\lambda_c)^{1/2} \sim \chi_F(\lambda_c)^{1/2} \sim N^{1/d \nu}.
\label{eq:fid_sus_scaling}
\end{equation} 
One may also look at $\chi_F(\lambda)$ away from the critical point where the expected scaling reads
\begin{equation}
\label{eq:fid_sus_away}
G(\lambda)^{1/2} \sim \chi_F(\lambda)^{1/2} \sim N^{1/2}|\lambda-\lambda_c|^{d\nu/2-1}.
 \end{equation}
The above universal contributions dominate the behavior of fidelity susceptibility when $d\nu<2$ so that non-universal, system-specific corrections remain subleading \cite{PolkovnikovArXiv2010,deGrandi2010a,*deGrandi2010b}. 

As a consequence, a realization of a physical system with small $\nu$ can lead to a hyper-sensitive  estimation of $\lambda$.   The standard and often considered, exactly solvable one-dimensional spin Ising chain where $\lambda$ corresponds to the transverse field exhibits the critical point with $\nu=1$, resulting in $\chi_F(\lambda_c)^{1/2}\sim N$ \cite{Chen2007,Invernizzi08,Damski2013,*Damski2014}. In the following we propose a physical realization of another spin system leading to much smaller value  of $d \nu < 1$. This provides a more intriguing example of a system which exhibits extreme sensitivity when $\lambda$ is varied across the critical point, and, on the first sight, might seem to break the Heisenberg limit.


\section{Metrology at the critical point}
\label{main}
In this section, we employ the adiabatic theorem to argue how slowly the parameter $\lambda$ has to change for the system to be 
able to adjust to it and follow instantaneous ground state. At the critical point this results in a time factor which scales as a power law with the system size. More generally, we show that the time dependence of QFI satisfies the bound where the time factorizes and the remaining scaling with $N$ can exceed the shot noise limit due to strong correlations in the critical ground state. It is however consistent with the Heisenberg limit  in Eq.~\eqref{eq:boixo}.  As such we reconcile this approach with the rotational scenario.  On the other hand, using finite size scaling analysis we argue that in the adiabatic regime the most natural observables, corresponding to part of the Hamiltonian coupled to $\lambda$, offer the same scaling of the error propagation formula as promised by the QFI.

\subsection{Characteristic time scale}
First, we estimate the  rate of changes of $\lambda$ which is needed for the system to stay in the instantaneous ground state.
We assume that 
\begin{equation}
 \delta_\lambda(t') = t'/\tau_Q = \frac{t'}{t} \delta_\lambda,
\label{eq:tauQ}
 \end{equation}
for $t' \in [0,t]$, where $t$ is the total time of the evolution, $\delta_\lambda(t) = \delta_\lambda$, and $\tau_Q =  t/\delta_\lambda$ is the quench rate. 
In order to estimate this rate we have to know the behavior of the energy gap at the critical point. For a continuous critical point this gap is expected to scale as
$\Delta E \sim L^{-z}$ which introduces the critical exponent $z$. We also need to estimate the width of the region of $\lambda$'s for which the gap is close to its minimal value. Standard finite size argument gives $\Gamma \sim L^{-1/ \nu}$. It follows from the general heuristics that in the finite system the gap would be comparable with its minimum when $L \sim \xi(\lambda) \sim |\lambda - \lambda_c|^{-\nu} = \Gamma^{-\nu}$. Now, the adiabatic conditions reads
$\Gamma \Delta E \gg 1/\tau_Q$, see e.g.~\cite{Knysh_2016}. This is equivalent to
\begin{equation}
\tau_Q \gg L^{\frac{z\nu +1}{\nu}}  = N^{\frac{z\nu +1}{d \nu}} .
\label{eq:adiabatic_KZ}
\end{equation}
The same estimate of the relevant quench rate is obtained from the application of  Kibble-Zurek argument \cite{Zurek2005,DziarmagaReview,polkovnikov_review2010}. The latter predicts the density of defects excited during the slow quench across the critical point. The adiabatic dynamics corresponds, in that case, to the extreme limit when no defects are created in a finite system.
 
 In order to induce the change of the parameter $\delta_\lambda \sim N^{-1/d \nu}$ which, according to Eq. \eqref{eq:fid_sus_scaling}, can possibly be observed, the time must scale at least as  
 \begin{equation}
\hat t \sim \tau_Q \delta_\lambda \sim N^{z/d}. 
 \label{eq:time_scale}
 \end{equation}
 Otherwise, Eq.~\eqref{eq:time_scale} simply represents the characteristic time scale at the critical point, given by the inverse of the energy gap in the finite system. This means that this time scale would naturally be relevant also beyond the scheme assuming adiabatic evolution and the ground-state overlap. We further elaborate on this point in Sec.~\ref{sec:le} where we briefly discuss how a small instantaneous quench and Loschmidt echo naturally fits into the general picture discussed here.
 
We can now bringing those scalings together. The bound in Eq. \eqref{eq:boixo} applied to the ground-state fidelity scenario would then read 
\begin{equation}
N^{-1/d \nu} \sim G(\lambda_c)^{-1/2} \ge 1/\hat t ||\hat H_1|| \sim N^{-z/d-1},
\label{eq:scaling_reconcile}
\end{equation}
note that $|| \hat H_1 || \sim N$ which corresponds to the usual
Heisenberg factor. In this reasoning we make a straightforward generalization of the argument of \cite{boixo07}, valid for time independent systems, to the time dependent adiabatic evolution.
In the scaling sense Eq.~\eqref{eq:scaling_reconcile} is equivalent to the condition that $(z+d) \nu \ge 1$. 

At this point it is convenient to introduce the scaling dimension of the operator $\hat h$, $[h]$, which describes the rescaling of the operator upon the scale transformation at the critical point. 
It gives the power-law behavior of the connected correlation function $C(r) = \langle \hat h_n \hat h_{n+r}\rangle - \langle \hat h_n \rangle \langle \hat h_{n+r} \rangle \sim r^{-2 [h]}$ in the thermodynamic limit.
The scaling exponents are not independent but, as we have one relevant operator here, can be typically expressed as a combination of $[h]$, $z$ and $d$. For instance $\nu = 1/(d+z - [h])$ \cite{CardyBook,ContinentinoBook}.
As $[h]\ge0$, this shows that Eq.~\eqref{eq:scaling_reconcile} is indeed consistent within our scaling discussion as $(z+d) \nu \ge 1$ holds. 

\subsection{Error propagation formula in the adiabatic limit}
Second, we focus on the adiabatic limit, where we discuss the scaling of the error propagation formula of the most natural observables $\hat H_1$ and $\hat h = \hat h_{N/2}$. We assume $\hat h$ to be in the bulk of the system to avoid possibly effects related with the boundaries of the system.
It is an exercise in finite size scaling analysis to argue that at the critical point 
\begin{eqnarray}
 \Delta_{\delta_\lambda}(\hat H_1,\lambda_c) & \sim & N^{-1/d \nu},  \label{eq:epfA1} \\
 \Delta_{\delta_\lambda}(\hat h_{N/2},\lambda_c) & \sim & N^{-1/d \nu + [h]/d}.
 \label{eq:epfA2}
 \end{eqnarray}
 The first one is saturating the bound provided by the fidelity susceptibility and the second one is close to it for small $[h]$  -- see below for the derivation under the assumption that in $d$-dimensional system  the correlation function is vanishing with distance $r$ slower than $r^{-d}$, and that the hyper-scaling relations hold. 
Those scalings are closely connected with the important observation that fidelity susceptibility (QFI) can be directly calculated by integrating the dynamic susceptibility of the system to the external driving $\hat H_1$ \cite{You2007,zanardi_geometric,Schwandt2009,ABQ2010,Hauke16}.

In order to derive Eqs.~(\ref{eq:epfA1}) and (\ref{eq:epfA2}) we analyze the scaling of the standard deviation and susceptibility appearing in the error propagation formula in Eq.~\eqref{eq:epf}. 
In the thermodynamic limit the susceptibility $\partial_\lambda \langle \hat h \rangle \sim |\lambda-\lambda_c|^{-\theta}$. We assume that the hyper-scaling law holds, i.e. that there are no dangerous irrelevant operators which could modify the scaling hypothesis. In that case $\theta =  1- [h] \nu$.
It is expected to hold for sufficiently low-dimensional system, below the so called upper critical dimension. This is the limit of interest from the perspective of quantum enhanced metrology, as the quantum effects in quantum many-body systems are becoming less important with the growing dimension of the system due to the monogamy of entanglement.
In the above, we also assume that $\theta \ge 0$. Otherwise non-universal effects dominate the behavior of susceptibility and effectively $\theta=0$.
Now, the standard finite size scaling argument implies that for a finite system at the critical point $\partial_\lambda \langle \hat h \rangle \sim (L^{-1/\nu})^{-\theta} \sim N^{\theta/d\nu }$. Assuming that the standard deviation ${\mathrm{std}} (\hat h) \sim 1$ leads to Eq.~\eqref{eq:epfA2}, where we have used the hyper-scaling relation. 

Similarly,  the susceptibility $\partial_\lambda  \langle \hat H_1 \rangle \sim N^{1+\theta /d\nu}$ follows from the scaling for $\hat h$ (times factor of $N$, we additionally assume that possible boundary effects are subleading). It is then enough to estimate the behavior of the standard deviation where we have to take into account the correlator $C(r) \sim r^{-2[h]}$ in the ground state at the critical point. The leading behavior is obtained by integrating the correlation function over the $d$-dimensional ball of radius $L$. If $C(r)$ is not vanishing faster than $r^{-d}$, i.e. for $d-2[h]>0$,  the integral is dominated by the tail of the correlation function and gives $\sqrt{\langle \hat H_1^2 \rangle - \langle \hat H_1 \rangle^2} \sim L ^ {d-[h]} = N^{1-[h]/d}$. 
Note that the standard deviation corresponds to a structure factor at $k=0$. Combining the expected scaling of standard deviation and susceptibility, together with the hyper-scaling relation gives Eq.~\eqref{eq:epfA1}.

It is worth to discuss the case of $d-2[h] \le0$ as well. Here, $\sqrt{\langle \hat H_1^2 \rangle - \langle \hat H_1 \rangle^2} \sim L ^ {d/2} = N^{1/2}$ and consequently $\Delta_{\delta_\lambda}(\hat H_1,\lambda_c) \sim N^{-1/d\nu + [h]/d -1/2}$. It does not saturate the bound given by QFI and we only see the classical $N^{-1/2}$ improvement over the single-site measurement in Eq.~\eqref{eq:epfA2}.
This is, for instance, the case in the often discussed quantum Ising spin chain in the transverse field, $\hat H = - \sum_{n=1}^N \sigma^x_n \sigma^x_{n+1} + g \sigma^z_n$. It has a critical point for $g_c=1$ with the exponent $z=1$. When $\hat h_n = \sigma^z_n$ corresponds to the transverse field the scaling dimension $[h] = 1$, $\nu=1$, and effectively $\theta=0$. The error propagation formula for $\hat H_1 = \sum_n \sigma^z_n$ was calculated in \cite{Kraus16} and does not saturate the scaling of $G^{-1/2}(g_c) \sim N$. It reads $\Delta_{\delta_\lambda}(\hat H_1, \lambda_c) \sim (N \log(N))^{-1/2}$ in agreement with the general prediction above. We note that logarithmic corrections to the scaling are typically expected in this case as $\theta=0$. 

\subsection{Consistency with the rotational scenario}
The above scalings of the error propagation formula assume adiabatic dynamics and as such would be recovered for long enough evolution times. At this point we bring the time explicitly into the picture. We show that
\begin{equation}
G(\lambda,t)^{1/2} \le t 2 \zeta \sqrt{\langle \hat H_1^2 \rangle - \langle \hat H_1 \rangle^2}.
\label{eq:Gt}
\end{equation}
where the standard deviation is calculated in the initial ground state. The factor $\zeta = \frac{1}{t \delta_\lambda} \int_0^{t} dt' \delta_\lambda(t')$ follows from the evolution profile in Eq.~\eqref{eq:tauQ}. Here $\zeta=1/2$.
Eq.~\eqref{eq:Gt} represents the quantum speed limit adjusted to our setting. For a recent review on quantum speed limits, see e.g. Ref.~\cite{Deffner_QSL_2017}.

For a system at the critical point, as discussed in the previous section, this leads to
\begin{equation}
G(\lambda_c,t)^{1/2} \lesssim t 2 \zeta N^{1-[h]/d}.
\label{eq:Gt_critical}
\end{equation}
We assume here that the correlations $C(r)\sim r^{-2[h]}$ do not vanish to quickly and $d>2[h]$. Obviously,  as $[h] \ge0$, Eq.~\eqref{eq:Gt_critical} is within the general bound (valid for any initial state) given by the Heisenberg limit.
In this article -- and more generally in the ground state fidelity approach -- the focus is on the ground state of the critical system and evolution generated by the Hamiltonian slightly perturbed from it. As can be seen in Eq.~\eqref{eq:Gt_critical} this allows to go beyond the shot noise limit, $G(\lambda,t)\sim t N^{1/2}$, as
a result of strong entanglement of such an initial state and algebraically vanishing $C(r)$. 

For the system detuned from criticality, or when $d-2[h] \le0$, the variance in Eq.~\eqref{eq:Gt} is not super-extensive and the classical scaling with $N$ is recovered. This would be again the case for the Ising model briefly discussed at the end of previous section. There, for $\hat h_n = \sigma^z_n$ corresponding to the transverse direction, $C(r)\sim r^{-2}$ at the critical point.

It is worth to compare this with the rotational scenario, in which case the suitable GHZ-type probe state is usually considered. We would then have $C(r)\sim1$ and effectively $[h] = 0$, which saturate the scaling of the Heisenberg limit (at least for short times). Interestingly, critical spin Hamiltonian for which GHZ-state is the ground state can be supplied \cite{comment_GHZ,Wolf_PRL_2006}.

The scalings in Eqs.~(\ref{eq:fid_sus_scaling},\ref{eq:epfA1},\ref{eq:epfA2}) are reached for the evolution times of the order of $\hat t \sim N^{z/d}$, Eq.~\eqref{eq:time_scale}. They comprise the ultimate limits of the criticality-based quantum metrology,  giving the title of this article a proper meaning.

To derive Eq.~\eqref{eq:Gt} we straightforwardly generalized the result of \cite{Pasquale2013,Dur,Pang_2014} to the time-dependent Hamiltonian, with $\delta_\lambda(t')$ e.g. as in Eq.~\eqref{eq:tauQ}. This leads to \cite{comment_Dur}
\begin{equation*}
G(\lambda,t) = 4 \zeta^2 t^2 \left( \langle \Psi | \hat O_1^2 | \Psi \rangle - \langle \Psi | \hat O_1| \Psi \rangle^2 \right) = 4 \zeta^2 t^2\mathrm{var}\left(\hat O_1\right),
\end{equation*}
where $\hat O_1 = \frac{1}{t\delta_\lambda \zeta} \int_0^t dt' \delta_\lambda(t') U(\lambda, t-t')^\dagger \hat H_1 U(\lambda, t-t')$ is time averaged operator $\hat H_1$ rotated by $U(\lambda, t) = e^{-i t \hat H(\lambda)}$ and $| \Psi \rangle$ is the probe state. As variance is convex, we have 
$G(\lambda,t)  \le 4 t^2 \zeta^2 \frac{1}{t \delta_\lambda \zeta} \int_0^t dt' \delta_\lambda(t')  \mathrm{var}\left( U(\lambda, t-t')^\dagger \hat H_1 U(\lambda, t-t')\right)$.
In our setup we consider the initial state which is the ground state of $\hat H(\lambda)$, $| \Psi \rangle = | \Psi(\lambda) \rangle$, which leads to Eq.~\eqref{eq:Gt}.

\section{Example: XXZ model in the external field}
\label{Sec_XYmodel}

The discussion in the previous section is general and should hold for a broad class of systems exhibiting continuous quantum critical points. In order to illustrate those predictions, in this section we consider the ferromagnetic XXZ spin-$\frac12$ spin chain in the external field. The Hamiltonian reads
\begin{equation}
\hat H(\lambda) = 
-\sum_{n=1}^{N-1}\left(\sigma^x_n\sigma^x_{n+1}+\sigma^y_n\sigma^y_{n+1} + J_z \sigma^z_n\sigma^z_{n+1}\right) +\lambda \sum_{n=1}^N  \sigma^x_n,
\label{eq:XYH}
\end{equation}
where we assume open boundary conditions. $N=L$ is the number of spins ($d=1$) and $J_z$ is an anisotropy parameter. 
We consider changes induced by the magnetic field $\lambda$ with other parameters fixed. For $|J_z| \le 1$ the system has a critical point at $\lambda_c = 0$ with the critical exponent $z=1$. The exponent $\nu$ was calculated in Ref.~\cite{Affleck99} and for fixed $-1<J_z<1$ reads 
\begin{equation}
\nu = \frac{2}{4-\arccos(J_z)/\pi},
\label{eq:nu_XXZ}
\end{equation}
which follows from the scaling dimension of the operator $\sigma^x$, $[{\sigma^x}] = \arccos(J_z)/2\pi$.
The desired condition of $d \nu <1$ is satisfied for all values of $|J_z|<1$. 

We note that fidelity susceptibility for quite similar XXZ model was studied e.g. in Refs.~\cite{Venuti2007,Sirker2010,Vekua2015} both at zero and non-zero temperature. There, however, the external magnetic field $\lambda$ was not present and the shift of parameters was induced by changing the value of $J_z$. This leads to a qualitatively different type of behavior related with Berezinskii–Kosterlitz–Thouless critical point and in that case the system does not exhibit super-extensive scaling of the fidelity susceptibility.

At the risk of multiplying the notation let us define the following observables 
\begin{equation}
\hat M_x \equiv \hat H_1= \sum_{n=1}^N \hat \sigma^x_n,
\label{eq:Mx}
\end{equation} 
which corresponds to the simultaneous measurement of magnetization on all sites, and 
\begin{equation}
\hat m_x \equiv \hat h_{N/2} = \hat \sigma^x_{N/2},
\label{eq:mx}
\end{equation} 
i.e. the on-site magnetization in the center of the system.
 
\begin{figure} [t]
\begin{center}
  \includegraphics[width= 0.95\columnwidth]{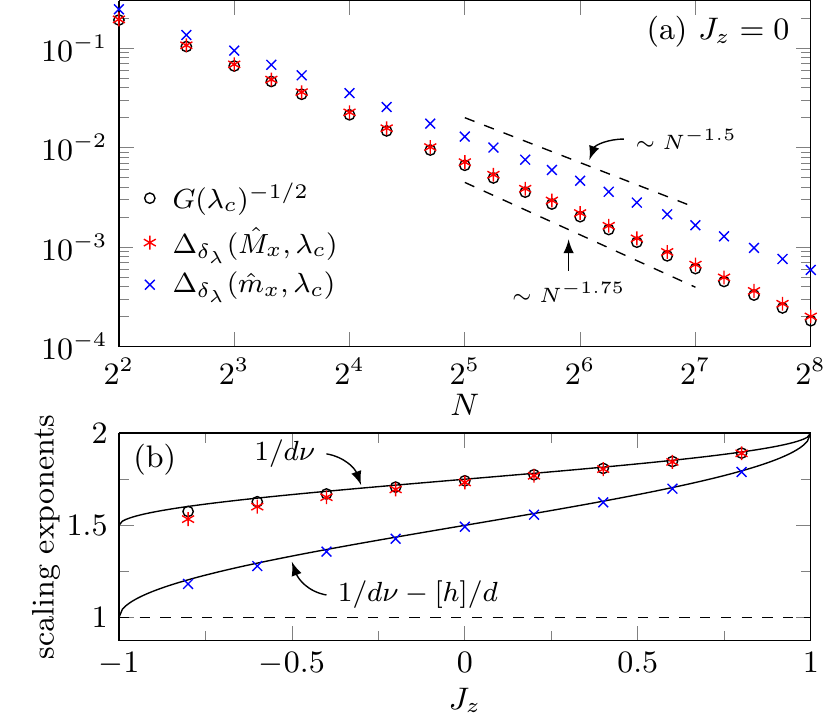}
\end{center}
  \caption{(Color online) XXZ model in the external field, Eq.~\eqref{eq:XYH}. (a) Scaling of the error propagation formula for operators $\hat M_x$ and $\hat m_x$, and the ultimate bound given by inverse of QFI, as a function of the system size.  The error propagation formula for $\hat M_x$ closely follows the ultimate bound.  Dashed lines indicate the slopes corresponding to the expected scaling and serve as guidance for an eye. The fits give $G(\lambda_c)^{-1/2} \sim N^{-1.74}$,  $\Delta_{\delta_\lambda}(\hat M_x,\lambda_c) \sim N^{-1.73}$ and $\Delta_{\delta_\lambda}(\hat m_x,\lambda_c) \sim N^{-1.49}$, where the expected exponents are $1.75$, $1.75$ and $1.5$, respectively. Here, $J_z = 0$ and the fits were done for $N=128 \div 256$ \cite{comment_evenN}.
 (b) In the considered model the scaling exponents in Eqs.~(\ref{eq:fid_sus_scaling},\ref{eq:epfA1},\ref{eq:epfA2}) depend continuously on the value of parameter $J_z$, following Eq.~\eqref{eq:nu_XXZ}. We compare those predictions with numerical results obtained similarly as in panel (a).  The exponent associated with the standard Heisenberg limit is marked with the dashed line for comparison. Its apparent breaking is the main subject of this article. }
   \label{fig:XXZ_scaling}
\end{figure}

\begin{figure} [t]
\begin{center}
  \includegraphics[width= 0.95\columnwidth]{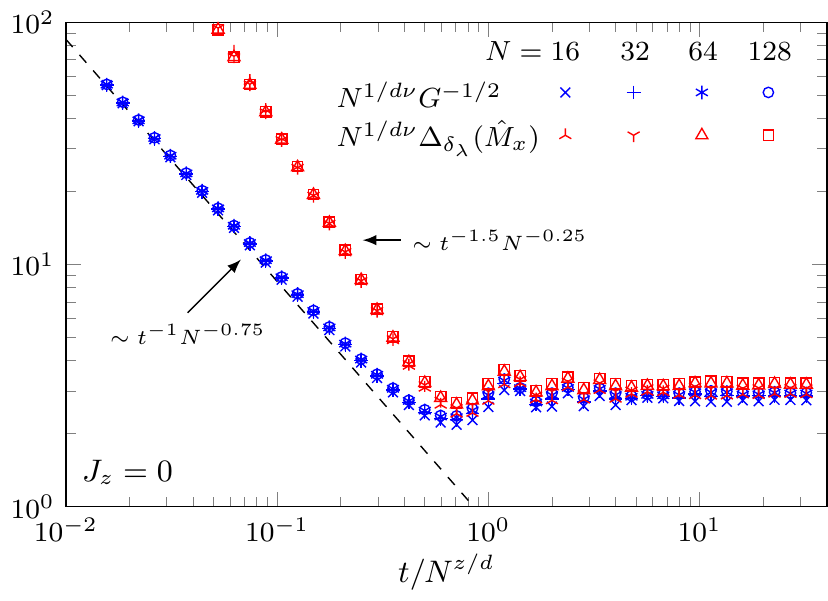}
\end{center}
  \caption{(Color online) Time dependence of QFI at the critical point in the XXZ model. Results from a small external magnetic field change as in Eq.~\eqref{eq:tauQ}.
   The time is rescaled by the system size according to \eqref{eq:time_scale}. QFI (blue symbols) is rescaled corresponding to the adiabatic limit given by Eq.~\eqref{eq:fid_sus_scaling}. Here, for $J_z = 0$,  $z/d = 1$ and  $d \nu = 4/7$. 
  For evolution times $t \gg N^{z/d}$ we recover the adiabatic limit. For short times $G^{1/2} \sim t N^{1-[h]/d}$ which is marked with dashed line. In our case $1-[h]/d = 3/4$. We also show the time dependence of the precision allowed by $\hat M_x$, as described by the error propagation formula (red symbols). For long time, in the adiabatic limit, it is almost optimal and nearly saturates QFI.
  For short times it is sub-optimal. In that case $\Delta_{\delta_\lambda}(\hat M_x,\lambda_c,t) \sim t^{-1.5} N^{-0.25}$ is below the shot noise limit as a function of $N$, see Eq.~\eqref{eq:DeltaM_time}, a result of strong fluctuation at the critical point. 
  } 
   \label{fig:XXZ_G_time}
\end{figure}

In Fig.~\ref{fig:XXZ_scaling} we calculate both QFI and the error propagation formula for $\hat M_x$ and $\hat m_x$ in the adiabatic limit. 
We numerically verify that the scaling relations in Eqs.~\eqref{eq:fid_sus_scaling}, \eqref{eq:epfA1} and \eqref{eq:epfA2} indeed hold in our model. Most importantly, it can be seen that the very natural  operator $\hat M_x$ practically reproduces the apparent super-Heisenberg scaling allowed by QFI. Moreover, while on-site magnetization in the bulk,  $\hat m_x$, grows slower with the system size then the ultimate bound, it is still well in the apparent super-Heisenberg regime.
For instance, Eqs.~(\ref{eq:epfA1},\ref{eq:epfA2},\ref{eq:nu_XXZ}) imply that for $J_z = 0$ we expect  $\Delta_{\delta_\lambda}(\hat M_x,\lambda_c) \sim N^{-7/4}$  and $\Delta_{\delta_\lambda}(\hat m_x,\lambda_c) \sim N^{-3/2}$. This is in excellent agreement with the numerical results presented in panel (a) of Fig.~\ref{fig:XXZ_scaling}. The exponents for other values of parameter $J_z$, both theoretical predictions and the values fitted from the numerics, are shown in panel (b).

The above results were obtained assuming that the evolution is adiabatic and the time of the evolution might have been, in principle, infinite.
We present the limitations imposed by finite evolution time in Fig.~\ref{fig:XXZ_G_time}. To that end the system was initialized in the ground state at the critical point $\lambda_c=0$. Subsequently, it was evolved to some infinitesimal $\delta_\lambda$
according to Eq.~\eqref{eq:tauQ}, with $\tau_Q = t/\delta_\lambda$ set by the total evolution time and $\delta_\lambda$. We then use Eq.~\eqref{eq:fidelity} to calculate QFI as a discrete derivative corresponding to Eq.~\eqref{eq:QFI_definition} (for small enough $\delta_\lambda$).

We rescale the evolution time with the characteristic time-scale in Eq.~\eqref{eq:time_scale} and the QFI according to the adiabatic (long time) limit in Eq.~\eqref{eq:fid_sus_scaling}.
As can be seen in Fig.~\ref{fig:XXZ_G_time} the rescaled data obtained for different system sizes collapse, corroborating the scaling predictions. QFI saturates at its adiabatic limit at the time given by Eq.~\eqref{eq:time_scale}. We obtain similar collapse for other values of the anisotropy parameter $J_z$ (not shown).

The bound on QFI given by Eq.~\eqref{eq:Gt} is plotted with the dashed line. The bound is tight in the limit of short times, in which case $G(\lambda_c,t)^{1/2} \sim t N^{3/4}$ for $J_z = 0$ plotted in Fig.~\ref{fig:XXZ_G_time}.
This scaling is obviously in full agreement with the Heisenberg limit. Employing the ground state of critical system as a probe allows, however, to go beyond the shot noise limit.

Finally,  in Fig.~\ref{fig:XXZ_G_time} we show how sensitivity allowed by $\hat M_x$ depends on time in our setup. As expected, for long enough times $\Delta_{\delta_\lambda}(\hat M_x,\lambda_c,t)$ saturates at the adiabatic value, almost saturating the ultimate QFI bound.
It is however far from being optimal for short times. In that case for $1\ll t \ll N^{z/d}$ we can expect \cite{comment_pertM} 
\begin{equation}
\Delta_{\delta_\lambda}(\hat H_1,\lambda_c,t) \sim t^{-\theta/z\nu} N^{-[h]/d}.
\label{eq:DeltaM_time}
\end{equation}
For $J_z = 0$ in Fig.~\ref{fig:XXZ_G_time} this translates into $\Delta_{\delta_\lambda}(\hat M_x,\lambda_c,t) \sim t^{-3/2} N^{-1/4}$. Such scaling of error propagation formula in the limit of short times follows from the behavior of susceptibility, i.e. denominator in Eq.~\eqref{eq:epf}.
The susceptibility $\partial_\lambda \langle \hat H_1 \rangle \sim N t^{\theta/z\nu}$ which directly follows from the universal scaling of dynamical susceptibility at the critical \cite{Sachdev,Shondireview}.
As the standard deviation $\sqrt{\langle \hat H_1^2 \rangle - \langle \hat H_1 \rangle^2} \sim N^{1-[h]/d}$ is set by the reference initial state, this gives Eq.~\eqref{eq:DeltaM_time}.
This derivation demonstrates that in this limit strong fluctuations -- corresponding to slowly vanishing correlation $C(r)$ -- limit the precision allowed by operator $\hat H_1$, putting it below the shot noise limit.
Conversely, we would be able to recover the short noise limit here for $C(r)$ vanishing faster then $r^{-d}$ (or away from the critical point).

All numerical results presented in this sections were obtained using the toolbox of Matrix Product States (MPS) \cite{verstraete2008matrix,schollwock2011,haegeman2011tdvp,*haegeman2014unifying}.
The time evolution is simulated using time dependent variational principle \cite{haegeman2011tdvp,*haegeman2014unifying}, which projects the Schr\"odinger equation onto the tangent space of the manifold of the MPS. For our problem we use the 4th order time-dependent Suzuki-Trotter decomposition \cite{Suzuki_decomposition_1993,*Hatano_review_2005},  necessary to split the unitary evolution operator onto parts acting on matrices of MPS associated with individual spins. We check that the results are converged both in the discreet time step and MPS bond dimension.

The natural question is how to produce or {emulate} the above physical system. Such a spin model may be possibly realized for repulsively interacting ultra-cold bosons in optical lattice potential in a quasi-one-dimensional geometry resulting from tight confinement in the perpendicular directions. 
The optical lattice potential can be precisely controlled, in particular, it can be shaken laterally 
\cite{Eckardt05,Eckardt10} that allows to change the system properties. By modulating intensity of laser 
beams forming the optical lattice its depth can be also modulated periodically \cite{Lacki13}.
Assuming that both processes occur with the same frequency $\omega$, frequency that is much larger
than the tunneling frequency as well as a characteristic frequency due to  interactions, one can 
derive an effective time-averaged Hamiltonian
governing the long-time physics as reviewed e.g. in \cite{Goldman14,Bukov2015,Eckardt2015}. Importantly, 
we assume, that this frequency (or rather its integer multiple, ${\cal N}\omega$) is 
resonant with $s \rightarrow p$ transition between the lowest $s$, and the excited, $p$, band. Such a resonance 
leads to additional slowly varying terms that affect the effective Hamiltonian obtained after time-averaging. 

Consider such a system with unit mean filling.   Identifying the proper ground 
state manifold one may describe the dynamics with an effective spin Hamiltonian.
Depending on ${\cal N}$ one may realize the effective XXZ Heisenberg model for ${\cal N}=2$, 
or  the model that reduces to the desired XXZ Heisenberg Hamiltonian in the magnetic field in Eq.~\eqref{eq:XYH} for ${\cal N}=3$. 
For the interested readers, we provide the explicit derivation in Section \ref{sec:cold} while we first show that analogical universal behavior holds for instantaneous quench of the Hamiltonian, i.e. the Loschmidt echo. We discuss as well the possible effects  due to an imperfect tuning and a finite temperature.

\section{Loschmidt echo}
\label{sec:le}

In the previous sections, in order to be able to recover the limit of adiabatic evolution, we were considering parameter $\lambda$ changing smoothly in time as in Eq.~\eqref{eq:tauQ}. For completeness of the discussion we briefly comment that qualitatively similar behavior is obtained in the other extreme limit, namely that of a sudden quench. Such a situation is closer to the original spirit of the rotational scenario where the Hamiltonian generating the evolution is usually time independent. To that end we are again going to initialize the system in the ground state of the initial Hamiltonian, focusing on the critical point as the most interesting regime, and consider evolution generated by the Hamiltonian detuned by $\delta_\lambda$. Fidelity in Eq.~\eqref{eq:fidelity} gives directly the so called Loschmidt echo which received significant attention in the literature \cite{DuttaBook}, for instance in the studies of decoherence. Most notably for us, it was shown that the decay of the Loschmidt echo is 
enhanced at the vicinity of the quantum critical point \cite{Loschmidt_Quan_2006}.

\begin{figure} [t]
\begin{center}
  \includegraphics[width= 0.95\columnwidth]{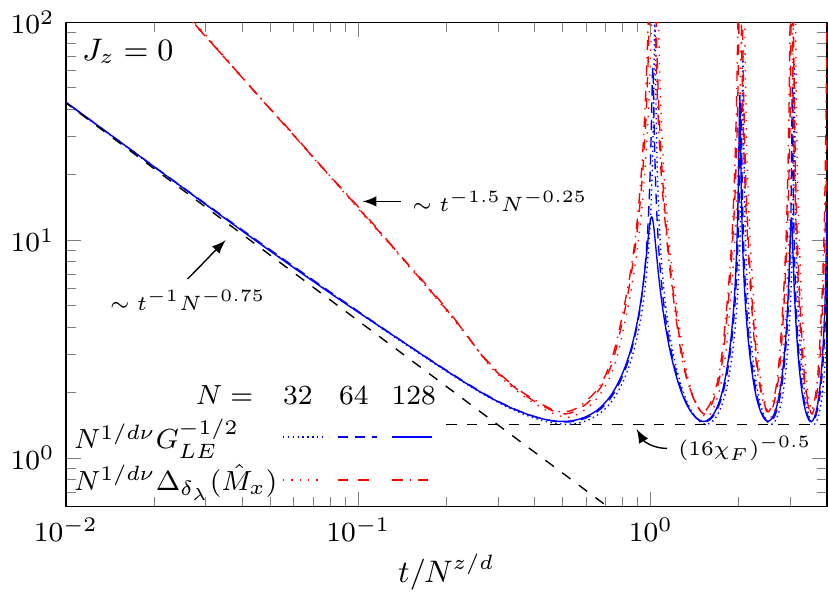}
\end{center}
  \caption{(Color online) Time dependence of QFI at the critical point in the XXZ model, which results from a small instantaneous shift of the external magnetic field, i.e. the Loschmidt echo.
   The time is rescaled by the system size according to Eq.~\eqref{eq:time_scale}. The maximal value which can be reached by QFI is bounded by the ground-state fidelity susceptibility (dashed line). We rescale QFI (blue lines) according to Eq.~\eqref{eq:fid_sus_scaling}.
   Here, for $J_z = 0$,  $z/d = 1$ and  $d \nu = 4/7$. For short times $G^{1/2} \sim t N^{1-[h]/d}$, marked with dashed line, with $1-[h]/d = 3/4$ in our case. For evolution time $t \sim N^{z/d}$ QFI almost reaches it maximal value and later exhibits periodic revivals characteristic for the Loschmidt echo.
   We also show the time dependence of the precision allowed by $\hat M_x$ (red lines). For long time it is close to optimal, mimicking the behavior of QFI. 
  For short times it is sub-optimal. In that case $\Delta_{\delta_\lambda}(\hat M_x,\lambda_c,t) \sim t^{-1.5} N^{-0.25}$ is below the shot noise limit as a function of $N$, see Eq.~\eqref{eq:DeltaM_time}. 
  } 
   \label{fig:XXZ_LE_time}
\end{figure}

We illustrate the time dependence of the QFI calculated in this setting, $G_{LE}(\lambda,t)$, for our XXZ model at the critical point with $J_z=0$ in Fig.~\ref{fig:XXZ_LE_time}. The scaling behavior is similar to the smooth quench shown in Fig.~\eqref{fig:XXZ_G_time}.
Indeed, the general bound in Eq.~\eqref{eq:Gt} directly applies to this case, where now $\delta_\lambda(t') = \delta_\lambda$ and $\zeta=1$.  For very short times the bound is tight \cite{Peres_LE_1984}, which follows from the perturbation theory \cite{comment_perturb}, and at the critical point we again get $G^{1/2}_{LE}(\lambda_c,t) \sim t N^{1-[h]/d}$ (under the assumption that $C(r)$ is not vanishing to quickly) \cite{Venuti_LE_2010,Dutta_LE_2012}.
As we consider evolution of the initial ground state by the Hamiltonian which is slightly detuned from the initial one, it is easy to see that QFI cannot grow unbounded and $G^{1/2}_{LE}(\lambda,t) \le 2 G^{1/2}(\lambda)$. Here $G(\lambda) = 4 \chi_F(\lambda)$ corresponds to ground state fidelity. At the critical point the bound is reached at the time scale given by Eq.~\eqref{eq:time_scale}.
In opposite to the smooth (adiabatic) scenario, Loschmidt echo displays revivals visible as peaks in Fig.~\ref{fig:XXZ_LE_time}, characteristic for the critical point \cite{DuttaBook}. The operator $\hat M_x \equiv \hat H_1$ again proves to offer a near optimal precision for long times. For short times the derivation in the previous section and Eq.~\eqref{eq:DeltaM_time} are expected to similarly hold, as can indeed be seen in Fig.~\ref{fig:XXZ_LE_time}.

\section{Robustness to detuning from criticality}
\label{sec:detune}

To complete the study of scaling results for the error propagation formula, it is quite natural  to ask to what extend such a scaling is relevant in real systems, for example due to a non-perfect tuning to the phase transition point and/or a finite temperature of an experiment. Fortunately, general scaling predictions addressing this issue can be provided and verified by numerical simulations, which we show in this section.

 First, when $\lambda$ is not tuned sufficiently close to the critical point, the fidelity susceptibility depends linearly on $N$ as discussed around Eq.~\eqref{eq:fid_sus_away}. The crossover is expected for $L / \xi \sim L |\lambda-\lambda_c|^{\nu} \sim 1$, where $\xi$ is the correlation length.
This  means that in order to obtain an apparent super-Heisenberg scaling, $\lambda$ should be tuned to the critical point within $|\lambda-\lambda_c| \ll L^{-1/\nu}$. This is also the range of $\delta_\lambda$ which can be observed in this case.

In the opposite limit of $|\lambda-\lambda_c| \gg L^{-1/\nu}$, away from the critical point, the error propagation formula are expected to scale as
\begin{eqnarray}
\Delta_{\delta_\lambda}(\hat H_1,\lambda) \sim N^{-1/2} |\lambda-\lambda_c|^{1-d \nu/2}, \label{eq:DeltaMaway} \\
\Delta_{\delta_\lambda}(\hat h,\lambda) \sim N^{0} |\lambda-\lambda_c|^{-\theta}. 
\end{eqnarray}
The derivation is analogical as in Sec.~\ref{main}. The susceptibility $\partial_\lambda \langle \hat h \rangle \sim |\lambda-\lambda_c|^{-\theta}$, together with 
${\mathrm{std}} (\hat h) \sim 1$, trivially gives the second of the above relations. 
For the first one, the standard scaling argument estimates the standard deviation (or a static structure factor) as $\sqrt{\langle \hat H_1^2 \rangle - \langle \hat H_1 \rangle^2} \sim N^{1/2} \xi ^{d/2-[h]}  \sim N^{1/2} |\lambda - \lambda_c| ^{-\nu d/2+\nu[h]} $, where the correlations are approximately algebraic $\sim r^{-2[h]}$ up to a distance given by the correlation length $\xi$.
Similarly as in Sec.~\ref{main}, we assume here that the algebraic part of the correlation function is vanishing slower then $r^{-d}$. Otherwise long-distance behavior contributes subleadingly to the standard deviation.  Finally, combining this with the susceptibility $\partial_\lambda \langle \hat H_1 \rangle\sim N |\lambda-\lambda_c|^{-\theta}$, together with the hyper-scaling relation for $\theta$, gives Eq.~\eqref{eq:DeltaMaway}.

\begin{figure} [t!]
\begin{center}
  \includegraphics[width= 0.95 \columnwidth]{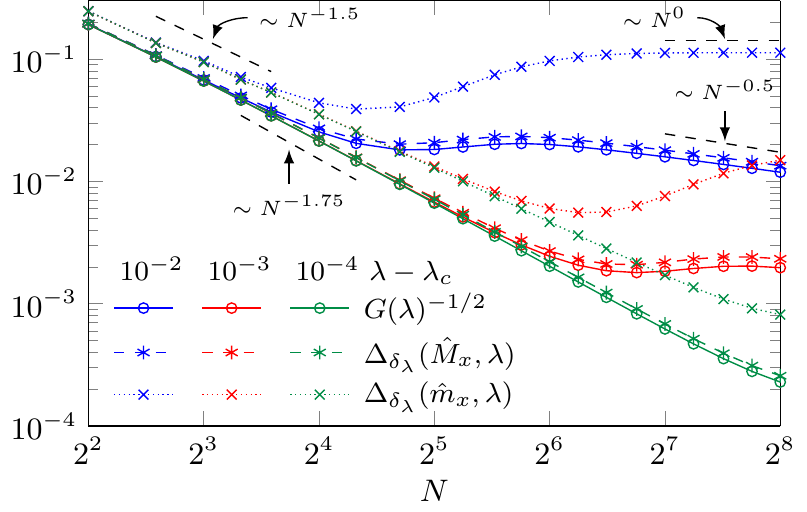}
\end{center}
  \caption{(Color online) The crossover between different scaling limits when $\lambda$ is not tuned exactly to the critical point. Results correspond to the adiabatic limit of the evolution.
  For given deviation $\lambda-\lambda_c$ we recover the apparent super-Heisenberg scaling when $N$ is small enough. When the system size is further increased and $N/\xi^d \sim N |\lambda-\lambda_c|^{d \nu} \gg 1$, $G(\lambda)^{-1/2}$ and $\Delta_{\delta_\lambda}(\hat M_x,\lambda)$ have a crossover to the classical $N^{-1/2}$ dependence. On the other hand, $\Delta_{\delta_\lambda}(\hat m_x,\lambda)$ saturates and becomes independent on the system size in that limit.  Nevertheless, notice that even in this case the prefactors in front of $N^{-1/2 (0)}$ are enhanced by the vicinity of the critical point.   $\Delta_{\delta_\lambda}(\hat M_x,\lambda)$  is closely following the optimal bound set by QFI for all values of the parameters.  Dashed lines indicate various scalings and serve as guidance for an eye. Results for the XXZ model in the external field in Eq.~\eqref{eq:XYH} with $J_z=0$.} 
   \label{fig:XXZ_detune}
\end{figure}

In Figure \ref{fig:XXZ_detune} we numerically verify those scaling predictions for $J_z=0$ in XXZ model in Eq.~\eqref{eq:XYH}.
While away from the critical point the accuracy allowed by $m_x \equiv \hat h = \sigma^x_{N/2}$ becomes significantly worse then the optimal one, the accuracy of $\hat M_x \equiv \hat H_1 = \sum_n \sigma^x_n$ closely follows the ultimate bound set by QFI. 
It is worth pointing out that even though we have classical $N^{-1/2}$ scaling in this limit, being in the vicinity of the critical point significantly improves the prefactor as in general, for $d \nu < 2$, $|\lambda - \lambda_c|^{1-d \nu/2} \ll 1$.
{Similarly as discussed in the previous sections for the critical point, this enhancement comes at a price of suitably longer evolution times. Here, however, the characteristic time scale $\hat t \sim \xi^{z} \sim |\lambda-\lambda_c|^{-z\nu}$ is independent of $N$.}

Second, the pure state idealization  discussed so far cannot be fully realized due to the external noise, including the thermal one. 
Here, for simplicity, we consider the situation where the temperature $T$ is finite but  $\lambda=\lambda_c$ is exactly tuned to the critical point. As the finite-size energy gap at the critical point scales as $L^{-z}$, one expects an apparent super-Heisenberg behavior to hold for $T \ll L^{-z}$. 

In the opposite limit of $T \gg L^{-z}$ we recover the classical behavior. We expect
\begin{eqnarray}
\Delta_{\delta_\lambda}(\hat H_1,\lambda_c,T) \sim N^{-1/2} T^{ (1- d \nu/2) /  z \nu}, \label{eq:DeltaMT}\\
\Delta_{\delta_\lambda}(\hat h,\lambda_c,T) \sim N^{0} T^{ -\theta / z \nu} \label{eq:DeltamT}. 
\end{eqnarray}
To that end, in order to simplify the analysis, we assume that there is no line of thermal phase transitions terminating at the quantum critical point which could alter the behavior and we employ simple scaling analysis, see e.g. \cite{ContinentinoBook}.  This is the case for our exemplary XXZ model and more broadly for one-dimensional systems.

In this case, the susceptibility $\partial_\lambda \langle \hat h \rangle \sim T^{-\theta/z \nu }$ leads to the second of the above relations. With the correlation length $\xi_T \sim T^{-1/z}$, the estimation of the standard deviation gives $\sqrt{\langle \hat H_1^2 \rangle - \langle \hat H_1 \rangle^2 }\sim N^{1/2} \xi_{T} ^ {-d/2+[h]} \sim N^{1/2} T^{-d/2z + [h]/z }$, where we again assume that the algebraic part of the correlation function is vanishing slower then $r^{-d}$. Similarly as in the previous case, $\partial_\lambda \langle \hat H_1 \rangle \sim N T^{-\theta/z \nu }$, together with the hyper-scaling relation results in Eq.~\eqref{eq:DeltaMT}.

\begin{figure} [t!]
\begin{center}
  \includegraphics[width=0.95 \columnwidth]{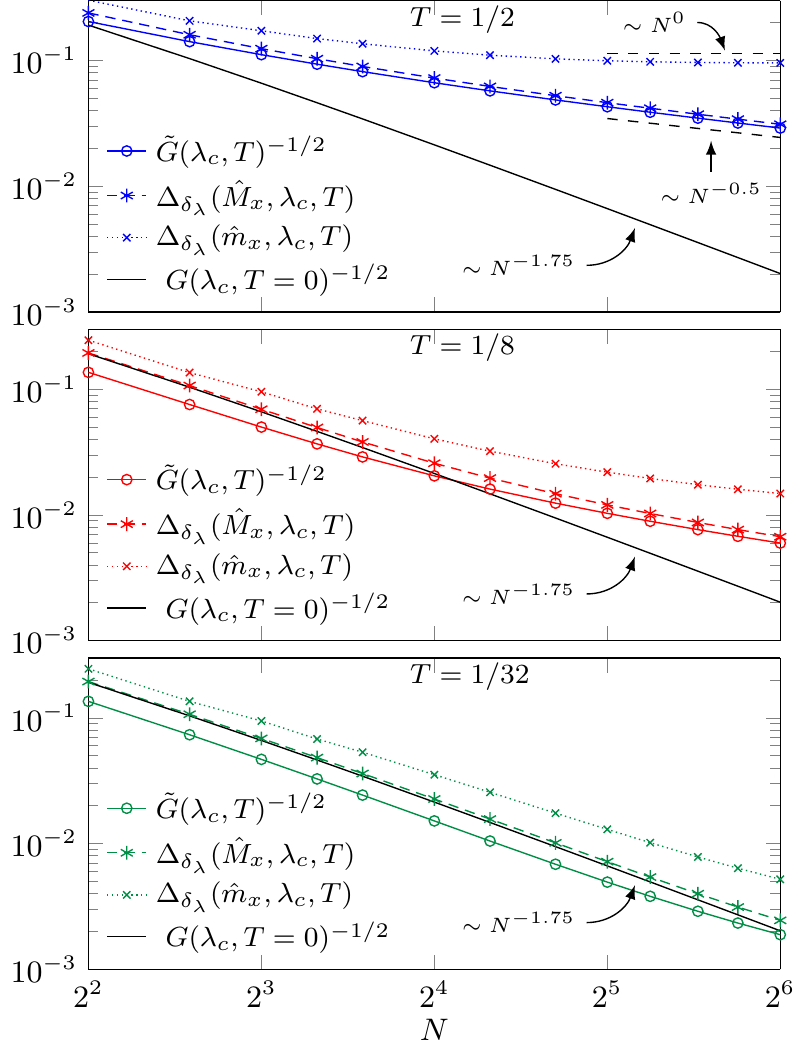}
\end{center}
  \caption{(Color online) The crossover between different scaling limits  when the temperature $T$ is non-zero. Results obtained by comparing states at equilibrium.
    For given small $T$ we recover the apparent super-Heisenberg scaling when $N$ is small enough,
  or alternatively for $T \ll N^{-z/d}$. In the opposite limit, when the system size is further increased and $N T^{d/z} \gg 1$, $\tilde G(\lambda_c,T)^{-1/2}$ and $\Delta_{\delta_\lambda}(\hat M_x,\lambda_c,T)$ have a crossover to the classical $N^{-1/2}$ dependence, however, with the prefactors that are enhanced by the presence of a quantum critical point at $T=0$. $\Delta_{\delta_\lambda}(\hat m_x,\lambda,T)$ saturates and becomes independent on the system size in that limit.  Similarly as in Fig.~\ref{fig:XXZ_detune}, we observe  that $\Delta_{\delta_\lambda}(\hat H_1,\lambda,T)$  is closely following the optimal bound set by QFI even at non-zero temperatures.  $\tilde G(\lambda_c,T)^{-1/2}$ plotted here is a lower bound of the (inverse of square root of) of the Fisher information, which lies between
  $\tilde G(\lambda_c,T)^{-1/2}$ and $(\tilde G(\lambda_c,T)/2)^{-1/2}$ -- see text for discussion. Solid lines show the corresponding ultimate bound for $T=0$ from Fig.~\ref{fig:XXZ_scaling}. Dashed lines indicate various scalings and serve as guidance for an eye. Results for $J_z=0$.    } 
   \label{fig:XXZ_T}
\end{figure}

We illustrate those scaling predictions in our model for $J_z=0$ in Fig.~\ref{fig:XXZ_T}. We employ MPS calculations, where the finite temperature density matrix $\hat \rho(\lambda) \sim e^{-\hat H(\lambda) / T}$ is expressed as purification and obtained via simulation of the finite system in the imaginary time \cite{verstraete2008matrix,schollwock2011,haegeman2011tdvp,haegeman2014unifying}. We again employ time dependent variational principle to that end.

Direct computation of fidelity in Eq.~\eqref{eq:fidelity} and, in particular, finding the positive square root appearing there is not feasible in the MPS representation.  We then follow Ref.~\cite{Sirker2010}
and in this case calculate the fidelity defined as $\tilde F(\hat \rho(\lambda), \hat \rho(\lambda + \delta_\lambda)) = \sqrt{{\mathrm Tr} \left[\hat \rho(\lambda)^{1/2} \hat \rho(\lambda + \delta_\lambda)^{1/2}\right] }$. Importantly, as discussed in \cite{Sirker2010} and derived in  \cite{ABQ2010} if one uses this definition to calculate fidelity susceptibility $\tilde \chi_F(\lambda)$ (similarly as in Eq.~\eqref{eq:rozwiniecie}) then $\tilde \chi_F(\lambda) \le \chi_F(\lambda) \le 2 \tilde \chi_F(\lambda)$. This allows us to define $\tilde G(\lambda,T) = 8 \tilde \chi_F(\lambda)$, which sets an upper bound on Fisher information, $\tilde G(\lambda,T)/2 < G(\lambda,T)  < \tilde G(\lambda,T)$, and which we plot in Fig.~\ref{fig:XXZ_T}. Those bounds cannot be tightened. Exact diagonalization done for systems of few spins suggests that 
$G(\lambda,T) \approx \tilde G(\lambda,T)/2$ in the limit of small enough $N$ or $T$. This is also seen in Fig.~\ref{fig:XXZ_T} from comparison to the data for $T=0$. In the opposite limit of large enough $N$ or $T$, $G(\lambda,T) \approx \tilde G(\lambda,T)$.

In our exemplary XXZ model $z=1$. Therefore, for $J_z =0$ and for a large enough system size or temperature, we expect the scalings $\Delta_{\delta_\lambda}(\hat H_1,\lambda_c,T) \sim N^{-1/2} T^{1.25}$ and $\Delta_{\delta_\lambda}(\hat h,\lambda_c,T) \sim N^{0} T^{1.5}$, which follow from Eqs.~(\ref{eq:DeltaMT},\ref{eq:DeltamT}). By fitting the temperature dependence to the numerical results for $N=2^6$, i.e. the largest size in Fig.~\ref{fig:XXZ_T}, and $T = 0.25 \div 1$ we obtain the exponents equal $1.2$ and $1.55$, respectively. This is in a reasonable agreement with the scaling predictions, especially given the numerical limitations. Simulations of the thermal states with MPS is typically much more demanding than the case of pure states, especially in the critical systems, which limits the system sizes which can be simulated here.

It is also worth pointing out that, as can be seen in Fig.~\ref{fig:XXZ_T}, $\Delta_{\delta_\lambda}(\hat H_1,\lambda_c,T)$ is again closely following the optimal bound set by an inverse of the Fisher information.  This is consistent with the scaling of QFI which,  similarly to Eqs.~(\ref{eq:fid_sus_scaling},\ref{eq:fid_sus_away}), can be deduced from the scaling dimension of the fidelity susceptibility \cite{zanardi_geometric,Schwandt2009} and standard scaling argument. Those give $\chi_F(\lambda_c,T) \sim N T^{(d\nu-2)/z\nu}$.

At the risk of stating the obvious, it is worth pointing out that the scaling predictions in Eqs. (\ref{eq:epfA1},\ref{eq:epfA2}) and in Eqs. (\ref{eq:DeltaMT},\ref{eq:DeltamT}) correspond to a different order of taking the limits of $T \to 0$ and $N \to \infty$ with the smooth crossover when the relevant order is changed.

Finally let us note that the finite temperature approach assumes implicitly a contract of the system with thermal reservoir, i.e., with thermal, Markovian noise. Thus the behavior observed is just the example of the classical scaling recovery for sufficiently large temperature  studied in detail recently for a more general case of an arbitrary Markovian noise \cite{Demkowicz17}.

\section{Cold-atom implementation}

\label{sec:cold}

In this section we discuss the possible implementation of the ferromagnetic Heisenberg Hamiltonian \eqref{eq:XYH}. We consider the
system of ultracold bosons trapped in the quasi 1D optical lattice 
subject to a periodic driving with frequency fulfilling the resonance condition \cite{Straeter15,Przysiezna15,Dutta15,Eckardt17}
 \begin{equation}
 \centering
 \mathcal{N} \omega = (E_p - E_s) + \mathfrak{d},
 \label{eq: tr_H007}
 \end{equation} 
 where $\mathfrak{d}$ is a small detuning from the transition between levels (bands)
 with energies $E_p$ and $E_s$. The dynamics of the system is captured within 
 the two--band Bose--Hubbard model 
\begin{equation}
 \centering
 \hat H_0 = - \sum_{\langle i,j \rangle}  \left( J_{s}  \hat s^{\dag}_i  \hat s_{j} + J_{p} \hat p^{\dag}_i  \hat p_{j} \right)
 + \sum_{i} \left( E_{s} \hat  n^{s}_i + E_{p} \hat n^{p}_i\right) + \hat H_{int},
 \nonumber
 \end{equation}
 where $J_s$ and $J_p$ are tunnelings in $s$ and $p$ bands, respectively, and the on--site repulsive interactions
are accounted for by
\begin{equation}
 \centering
 \hat H_{int} = \sum_{i} \left[  \frac{U_{ss}}{2}  \hat  n^{s}_i \left( \hat  n^{s}_i - 1 \right)
 +   \frac{U_{pp}}{2}  \hat  n^{p}_i \left( \hat  n^{p}_i - 1 \right) + \frac{U_{sp}}{2}   \hat  n^{s}_i \hat n^{p}_i \right].
 \end{equation}

The horizontal lattice shaking modifies the Hamiltonian by the term 
  \begin{equation}
 \centering 
 \hat H_{hor}(t) = \cos(\omega t) K  \sum_{i}   i \left( \hat n^{s}_i + \hat n^{p}_i \right) +
\nonumber
\vspace{-0.3cm}
 \end{equation} 
 \begin{equation}
   + J \cos(\omega t) \sum_{\langle i,j \rangle} \hat p_i^{\dag} \hat s_j + 
 W \cos(\omega t) \sum_i \hat p_i^{\dag} \hat s_{i} + h.c.,
 \end{equation}
where the constants $K, \, J, \, W$ depend on the amplitude of the periodic driving.  The modulation 
of the intensity of the laser field forming the optical lattice causes the on--site energies to oscillate
with amplitudes $\Delta_s$ and $\Delta_p$ at frequency $\omega_{v}$.
\begin{equation}
 \centering 
  \hat H_{ver}(t)=  \cos(\omega_{v} t)\sum_i \left(  \Delta_s \hat n^s_i+\Delta_p \hat n^p_i \right).
 \end{equation}
 We shall assume $\omega_{v}=\omega$ in the following for simplicity.
 The long--time dynamics of the system $H_0 + H_{hor}(t)+ H_{ver}(t)$ with $\mathcal{N} = 3$ is described
 by the effective time--averaged Hamiltonian
 \begin{equation}
 \centering 
\hat  H_{eff} =   \sum_{i } 
\left( J^+_{sp}( \hat p^{\dag}_i \hat s_{i+1}+\hat p_i \hat s^{\dag}_{i+1} ) +
J^-_{sp}(\hat  p^{\dag}_{i+1} \hat s_{i} + \hat p_{i+1} \hat s^{\dag}_{i})  \right)
\nonumber
\vspace{-0.2cm}
   \label{eq: effHam0010}
\end{equation}
  \begin{equation}
 \centering 
+  \sum_{\langle i,j \rangle}  \left( J^{\mathrm{ren}}_{s}  \hat s^{\dag}_i  \hat s_{j} + J^{\mathrm{ren} }_{p} \hat  p^{\dag}_i  \hat p_{j} \right) + 
W_{sp} \sum_i (\hat p^{\dag}_i \hat s_{i} + \hat p_i \hat s_{i}^{\dag}) + \hat H_{int}.
 \end{equation}
 The intra--band tunneling amplitudes are effectively $J^{\mathrm{ren}}_{s,p} = \mathcal{J}_0(K/\omega) J_{s,p}$, where $\mathcal{J}_0(x)$ is a 0-order Bessel function \cite{Eckardt05,Eckardt2015,Eckardt17}. Similarly, the shaking induced 
 inter--band tunnelings are renormalized by factors dependent on higher order Bessel functions yielding $J^{\pm}_{sp}$
and  $W_{sp}$ \cite{Sierant16}. Finally, the energies of the $s$ and $p$ states differ only by the detuning $ \mathfrak{d}$.  It is now assumed that $K/\omega = x_0 +  \epsilon$ (with $|\epsilon| \ll x_0$)
where $x_0$ is the first zero of $\mathcal{J}_0$ so that the hopping within the $s$ and $p$ bands is strongly suppressed.

 We are interested in the physics of excitations close to the ground state of Eq.~\eqref{eq: effHam0010} with the unit filling in the 
strongly interacting regime. The Hamiltonian $\hat H_{eff}$, within the second order of perturbation calculus 
becomes
$\hat H = - \hat  P_{g} \hat H_{eff} \left( \hat P_{e} \hat  H_{eff} \hat P_{e} - E  \right)^{-1} \hat H_{eff} \hat P_{g}$,
where $\hat P_{g}$ projects on the subspace of singly occupied states and $\hat P_{e} = 1 - \hat P_{g}$. 
The condition $\hat n^s_i + \hat n^p_i = 1$, which holds in the low energy subspace, enables one to define a spin $1/2$ degree of freedom 
at each lattice site leading to Hamiltonian \eqref{eq:XYH} with $\lambda \propto W_{sp} \propto \left((\Delta_p - \Delta_s)/\omega \right)^2  $ and parameter  $J_z$ depending on the values of $\epsilon$ and  $ \mathfrak{d}$.
Then the spin system in Eq.~\eqref{eq:XYH} effectively describes the excitations in the Mott space of the both laterally and vertically shaken optical lattice.

\section{Conclusions}
\label{sec:discuss}

Let us summarize our findings. We have discussed two approaches to quantum estimation of a parameter, approaches that give seemingly different predictions. This difference becomes apparent close to the critical phase transition points. In the first approach, referred to as a rotation scenario, the system is compared with its copy rotated by the parameter dependent dynamics. The ultimate limit in this case is known as the Heisenberg limit. The second approach relies on the overlap of ground states of the system at slightly different values of the parameter. Here one may often arrive at an apparent super-Heisenberg scaling {close to criticality}.

The main result of our work is to provide an unified picture that links these two approaches. The necessary ingredient is an observation that the physical comparison of ground states at different parameter values can be operationally realized in an adiabatic evolution only.  Under this assumption we provide {an argument} that both approaches yield essentially the same 
scaling, consistent with the Heisenberg limit when this time factor is taken into account. 
In effects we obtain a straightforward generalization of the argument of \cite{boixo07,Giovanetti06,Pasquale2013,Dur,Pang_2014}, valid for time independent systems, to an adiabatic evolution.

Importantly, our finding should hold for a broad class of a quantum many-body systems
(in one or more spacial dimensions), with a particular focus on the second-order quantum critical points. 
As a by product we identify the optimal observable that reveals the optimal scaling in the adiabatic limit - it is identified as a part of the Hamiltonian coupled to the parameter (i.e. $H_1$ in \eqref{ciacho}). 

The general result has been confirmed in a detailed study of the ferromagnetic Heisenberg Hamiltonian. On one side it forms a ``minimal'' Hamiltonian that reveals an apparent super-Heisenberg scaling with global magnetization as the optimal observable.  We have shown that this ``super-Heisenberg'' behavior is quite robust with respect to detuning from the critical point as well as temperature.  Yet, as verified in our numerical study, the time needed for measurement of the overlap (i.e. performing the necessary time evolution) leads to the recovery of the Heisenberg scaling in total agreement with the rotation scenario.

The standard metrological approach claims that
the scaling may grow with the range of the interaction involving 
the unknown parameter \cite{RoyBraunstein08},
i.e. super-Heisenberg behavior is in general possible
if we replace the one-body operators in Eq.~\eqref{ciacho} by multiple  
many-body terms. However, that may reduce the possible gap in the system and therefore affect the time needed to physically realize ground states 
the fidelity of which is supposed to be measured. The viewpoint developed here is that any super-Heisenberg claim must be accompanied by a
careful analysis not only of system size scaling but also the time needed to prepare a given measurement.

Specifically, cold atomic systems offer direct measurements of the fidelity susceptibility {\it without necessity of the unitary rotation} by means of the 
Bragg spectroscopy. That may allow to break the ``unitary rotation paradigm'' although the careful analysis of a specific experiment is needed before giving the definite answer.  In particular, high frequency resolution in Bragg spectroscopy and importance of low frequencies for fidelity susceptibility necessitates  a sufficiently long time of measurement. The discussion of that point is beyond the scope of the present paper. Similar remarks may be addressed to ``swap measurement''  (see the appendix A where this idea is developed) that also does not involve ``unitary rotation''. So while we have not provided all the answers leaving some place for future investigations, we believe that we were able at least to understand the apparent discrepancy between the unitary rotation approach and fidelity susceptibility behavior at criticality.

\begin{acknowledgments} We are grateful to Bogdan Damski and Jacek Dziarmaga for numerous discussions and hints.
P. S. and J. Z. acknowledge support by PL-Grid Infrastructure and EU project the EU H2020-FETPROACT-2014 Project QUIC No.641122, while
P. H. acknowledges the support of the ERC AdG QOLAPS project. This research has been supported by 
National Science Centre (Poland) under projects  2011/01/B/ST2/05459 (P.H.), 2015/19/B/ST2/01028 (P.S.), 2016/21/B/ST2/01086 (J.Z.) and 2013/09/B/ST3/00239 (M.M.R.). 

\end{acknowledgments}

\appendix
\section{An universal quadratic estimator for pure states}
\label{sec:swap}
Consider an arbitrary Hamiltonian $\hat H(\lambda)$ and suppose that we 
have significant reasons to believe that $\lambda$ is critical, but for technical 
reasons it is difficult to prove it. In particular, there is no linear 
estimator  known -- like the two observables discussed in the main text --
that could provide an accuracy close to the limit of  Fisher information scaling.
The question is whether there is any way to design an experiment that would allow to circumvent the above difficulty. 
To answer this affirmatively, we shall provide a simple estimator, quadratic in terms of 
the interaction involved, which however remains relatively simple and, at least 
in principle, can be detected with the present state of the art technology of optical lattices.

Let us assume that apart from $|\Psi (\lambda + \delta_\lambda) \rangle$, experimentalist can also prepare the state at the critical point  
$|\Psi (\lambda) \rangle $. Now, it is known that ${\cal F}^{2}$ -- the square of fidelity in Eq.~\eqref{eq:def_F} between the two above states -- 
 can be detected using the so called 
  universal quantum estimator \cite{Ekert-Et-Al-02,PHAE2002}, which is measurable involving at most quadratic interaction among elementary qubits corresponding to the series of independent Hong-Ou-Mandel
type measurements \cite{Miszczak09}.
The main idea behind it is to measure the quantity ${\mathrm Tr}(\hat \rho \hat \sigma)$, which is 
the mean value of the  {\it swap} observable (defined below) jointly measured
on  the product  state $\hat \rho \otimes \hat \sigma$.  
The first experiment of this type has been performed relatively long time ago on two copies of the same state of the
polarization-entangled photon pairs, which aimed at estimation of Renyi-2 entropy to show that violation of suitable inequality can serve as entanglement detector \cite{Bovino05}.

Consider the {\it swap} operator $\hat S$ which by definition acts as $\hat S|i\rangle |j\rangle = |j\rangle|i\rangle$ on ${\cal H} \otimes {\cal H}$.  
Alternatively, it can be represented as 
\begin{equation}
\hat S= \hat {\mathbb{1}}_{\cal H} \otimes \hat {\mathbb{1}}_{\cal H} - 2 \hat P^{asym}{}_{{\cal H} \otimes {\cal H}},
\end{equation}
where $\hat P^{asym}$ is the projector on the antisymmetric subspace of ${\cal H} \otimes {\cal H}$. Then the obvious observable which 
measures ${\cal F}^2$ is 
\begin{equation}
\hat A_{swap}=\hat S^{\otimes N}= \hat {\mathbb{1}}_{{\cal H}^{\otimes N}} \otimes \hat {\mathbb{1}}_{{\cal H}^{\otimes N}} - 2 \hat P^{asym}_{{\cal H}^{\otimes N} \otimes {\cal H}^{\otimes N}}
\label{eq:swap}
\end{equation}
acting on the ``quadratic'' state  
$|\Phi \rangle = |\Psi(\lambda) \rangle \otimes |\Psi(\lambda +\delta_\lambda)\rangle $ composed of $2N$  elementary subsystems, where
each $\hat S$ acts on one element of, respectively,  first and second pure states. 
The above measurement corresponds to coupling each elementary subsystem of the 
first chain of $N$ spins, with the corresponding subsystem of the second chain
and performing the measurement by projecting it on the antisymmetric subspace and then multiply the results.
This gives the overall results of $+1$ ($-1$) when the number of successive projections is even (odd).

Now it is an elementary exercise to see that in this case the error propagation formula of Eq.~\eqref{eq:epf} takes the form
\begin{equation}
 \Delta_{\delta_\lambda} (\hat A_{swap}, \lambda)  =  \frac{1}{\sqrt{2  \chi_{F}(\lambda)}}= \sqrt{\frac{2}{G(\lambda)}},
\label{error-swap}
\end{equation}
where we employ Eq.~\eqref{eq:rozwiniecie} and $\hat A_{swap}^2 = \hat {\mathbb{1}}$. 
The error propagation formula in Eq.~\eqref{error-swap} is then only by a factor of two worse then the best possible linear estimator.  Therefore, we got a quadratic estimator that reproduces -- up to the constant factor --  the  scaling in $N$ of the best possible linear one. 
Note that applying the above procedure to the original unitary perturbation scheme (like in eg. 
\cite{Demkowicz-review,Hauke16,Kraus16}) can not surpass the Heisenberg limit, since the whole scheme may 
be simulated  as a measurement of some new observable on the $|\Psi(\lambda)\rangle$ alone which is known to obey the 
limit as long as  the part of the Hamiltonian with unknown parameter $\lambda$ is fully local, see \cite{Pasquale2013,Dur, Fraisse2016}. 

Now the question is: can we employ the above approach for the case of bosonic lattices considered in Sec.~\ref{sec:cold}?
Fortunately, the answer is, at least in principle, affirmative.  Indeed, very recently it has been shown how to directly perform the measurement of the swap observable
on the bosonic lattices \cite{Daley12, Pichler13,Elben18,Vermersch18} where the authors designed the beam-splitter type of interaction as a tunneling coupling 
between two optical lattices, followed by the measurement of the parity of the on-site occupation numbers. 
This guarantees that the observable in Eq.~\eqref{eq:swap}, which is crucial for the experimental application of the formula \eqref{error-swap} can be directly measured 
on the optical system with an effective spin Hamiltonian having the critical parameter introduced by lattice shaking.


\begin{thebibliography}{111}
\bibitem[{com({\natexlab{a}})}]{comment_GHZ}
\bibinfo{note}{Following \cite{Wolf_PRL_2006} consider the cluster-Ising spin
  chain $H_0 = -\sum_{n} (2 \sigma^z_n \sigma^z_{n+1} + \sigma^x_n - \sigma^z_n
  \sigma^x_{n+1} \sigma^z_{n+2} )$ which is critical and has GHZ state as the
  ground state. Now, the relevant perturbation aligned with this state is $H_1
  = -\sum_n{h_n} = -\sum_n{\sigma^z_n}$. In such system $z=2$ and $[h]=0$.}
\bibitem[{com({\natexlab{b}})}]{comment_Dur}
\bibinfo{note}{See for instance Eqs.~(7-9) in Ref.~\cite{Dur}. In Eq.~(8) we
  allow a small change of, say, $\lambda_1$ to depend on time.}
\bibitem[{com({\natexlab{c}})}]{comment_evenN}
\bibinfo{note}{Experienced reader may wonder that for odd $N$ the ground state
  is exactly degenerate at the critical point $\lambda_c=0$, which leads to
  singularity in $\chi_F$ if one uses Eq.~\eqref{eq:def_F} directly. For
  simplicity of the discussion, we use even $N$ in the numerics.}
\bibitem[{com({\natexlab{d}})}]{comment_pertM}
\bibinfo{note}{In the perturbative regime of $t\ll 1$ it is easy to see that
  $\Delta_{\delta_\lambda}(\hat H_1,\lambda_c,t) \sim t^{-2}$. Such region is
  not shown in Fig.~\ref{fig:XXZ_G_time}.}
\bibitem[{com({\natexlab{e}})}]{comment_perturb}
\bibinfo{note}{Note that here the scaling remains tight also beyond the
  perturbative regime of very short times.}
\end{thebibliography}

\end{document}